\documentclass{iau} 
\usepackage{graphicx}

\title[JD 11.~~Pulsar Emission Physics]
{Pulsar Emission Physics: \\The First Fifty Years}

\author[Alice Harding]   
{Alice K. Harding$^1$}

\affiliation{$^1$ NASA Goddard Space Flight Center, Astrophysics Science Division, \\Greenbelt, MD 20771, USA \\ email: {\tt Alice.K.Harding@nasa.gov}}

\pubyear{2017}
\volume{337}  
\jname{Pulsar Astrophysics -­ The Next 50 Years}
\editors{P. Weltevrede, B.B.P. Perera, L. Levin Preston \& S. Sanidas, eds.}

\begin{document}

\maketitle

\begin{abstract}
Over the last fifty years since the discovery of pulsars, our understanding of where and how pulsars emit the radiation 
we observe has undergone significant revision.  The location and mechanisms of high-energy radiation are intimately tied to the 
sites of particle acceleration. The evolution of emission models has paralleled the development of increasingly more 
sensitive telescopes, especially at high energies.  I will review the history of pulsar emission modeling, from the early 
days of gaps at the polar caps, to outer gaps and slot gaps in the outer magnetosphere, to the present era of global 
magnetosphere simulations that locate most acceleration and high-energy emission in the current sheets. 
\end{abstract}

\firstsection 
\section{Introduction}

It is impossible to give a full review of the history of pulsar emission physics over the last fifty years in six pages.  I will only review what I consider the highlights and major influential advances to give an overall account of the development of ideas and their interaction with observations.  This review also concentrates on high-energy emission models since it is more strongly coupled to the overall pulsar energetics and global field structure than the radio emission.  More comprehensive recent reviews by \cite{Beskin2017}, \cite{CeruttiBeloborodov2016} and \cite{GrenierHarding2015} contain more detailed discussion.  In addition, \cite{Ferrara2017}  reviews the multiwavelength observations of pulsars.  

Figure 1 gives a historical timeline of the major models and ideas over the last fifty years.  Radiation models, that aim to describe the observations, and magnetosphere models, that aim to describe the global distribution of fields, currents and particles, developed along two separate tracks until recently.   These two tracks began to interact shortly after 2000, when advances in computing power finally allowed the development of realistic global magnetosphere models.  In the past few years, the tracks have merged as the global models finally provided self-consistent currents with regions of accelerating electric fields, enabling radiation modeling.  Also shown in Figure 1 are the major $\gamma$-ray telescopes that played an important role in the development of models.  The period from the mid-1980's to around 2000 marked a time of few important advances or major ideas, the ``Dark Ages" of pulsar physics (\cite[Beskin 2017)], which interestingly coincides with a long drought in data from $\gamma$-ray telescopes between the launch of COS-B in 1975 to the Compton Gamma-Ray Telescope in 1991.  The launch of the Fermi Gamma-Ray Space Telescope in 2008 drove a renaissance in model development that continues to the present.  

\begin{figure}[t]
\vspace*{-1.0 cm}
\begin{center}
 \includegraphics[width=3.5in,angle=90]{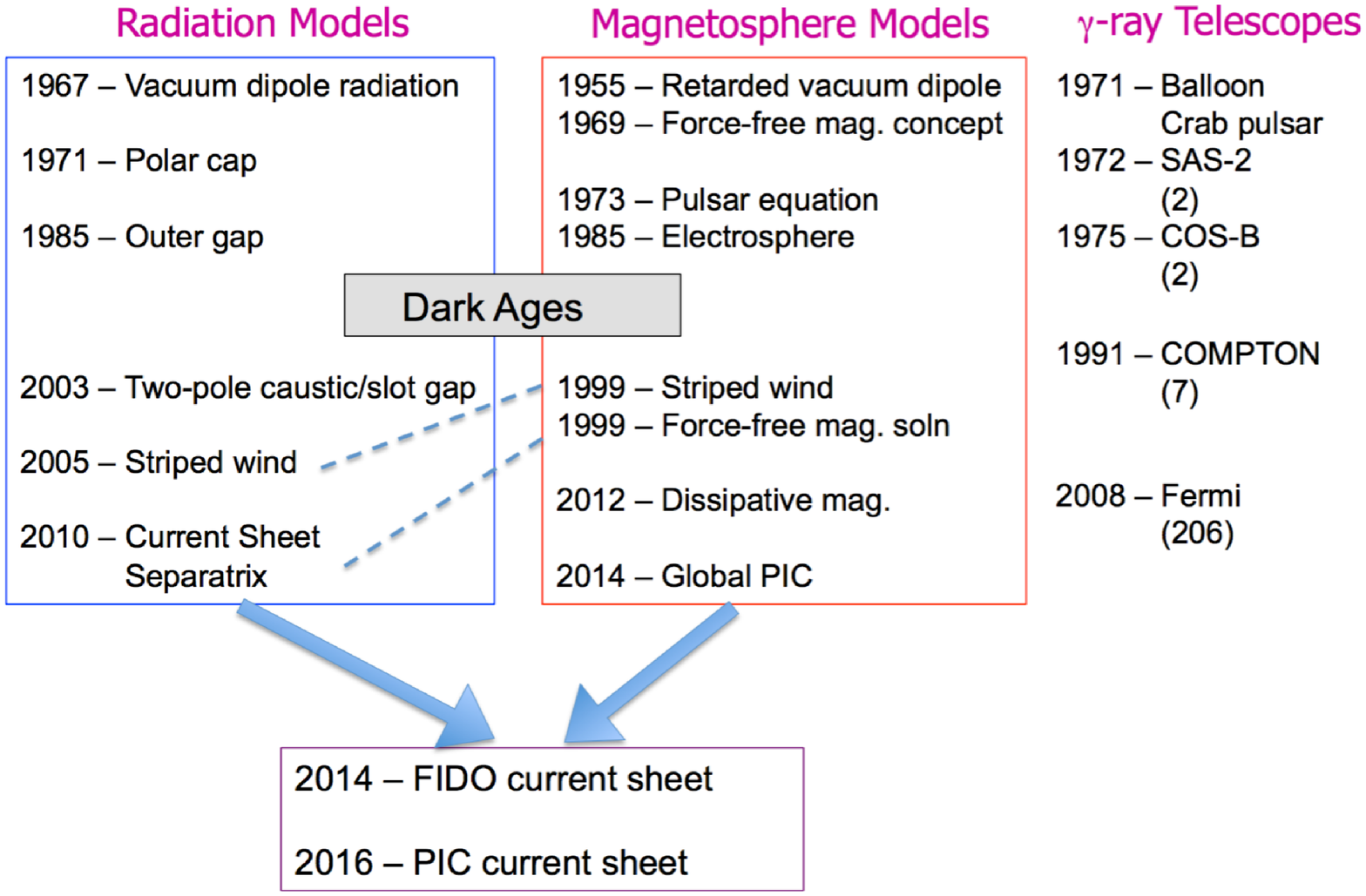} 
\vspace*{-1.0 cm}
 \caption{A timeline of theoretical developments in pulsar physics over the last 50 years.}
   \label{fig1}
\end{center}
\end{figure}

\section{Radiation Models}

Well before pulsars were discovered, \cite{BaadeZwicky1934} proposed that very compact stars supported by neutron degeneracy pressure could result from a supernova core collapse.  Very early models following pulsar discovery indeed focussed on neutron stars as being the only objects that could rotate so rapidly without breakup and could account for the precisely predictable radio pulse arrival times (\cite[Gold 1968)], leading to the ``lighthouse" model of radio pulsations.  These models also assumed that the star was surrounded by a vacuum, and the energy loss rate from a rotating magnetic dipole in vacuum matched well with the power radiated by the Crab nebula (\cite[Pacini 1968)]).  \cite{OG1969} further developed the rotating neutron star model, showing that the predicted spin-down luminosity $\dot E$ matched the available observations, and that implied huge surface magnetic fields around $10^{12}$ G and characteristic ages $\tau \sim P/\dot P \sim 10^3 - 10^6$ yr, where $P$ and $\dot P$ are the pulsar rotation period and  period derivative.  The solution for a magnetized, conducting star had in fact been published by Deutsch (1955) and showed that the dipole field lines are swept back near and beyond the (LC), $R_{\rm LC} = c/\Omega$ where $\Omega = 2\pi/P$, into a toroidal pattern that becomes an electromagnetic wave at infinity.  However, \cite{GoldreichJulian1969} in a landmark paper noted that the neutron star cannot be surrounded by vacuum since the large surface electric field induced by the rotation, ${\bf E = (v \times B)/c}$, far exceeds the gravitational force, so that particles will be pulled out of the surface to fill the magnetosphere with charges.  They defined the charge density in the magnetosphere needed to short out the electric field, $E_\parallel$, parallel to the magnetic field, the now well-known Goldreich-Julian (GJ) charge density, $\rho_{\rm GJ} = {\bf (\nabla \cdot E)}/4\pi \simeq -{\bf (\Omega \cdot B)} / 2\pi c$.  

The first models thus located the particle acceleration at the polar caps, with \cite{Sturrock1971} pointing out that accelerated electrons could radiate curvature radiation (CR) photons at $\gamma$-ray energies that would produce electron-positron pairs by the QED process of one-photon pair production (\cite[Erber 1966)].  However, if $\Omega \cdot B < 1$, the GJ charge density is positive above the polar caps but the required ions would be trapped due to the high surface work function (\cite[Ruderman \& Sutherland 1975)].  The resulting vacuum gap can break down through production of $e^+-e^-$ pairs by $\gamma$-rays entering the gap and an ensuring cascade.  After breakdown, the $E_\parallel$ builds up again producing another cyclic pair cascade and breakdown.  \cite{RudermanSutherland1975} suggested that the bursts of pairs that drift with velocity $E \times B$ around the polar cap could account for the observed drifting sub-pulses and that the counter streaming pairs would produce coherent CR.  Newer calculations of the neutron star surface work function for electrons and ions \cite{MedinLai2007}, though, found that for all but very high (magnetar-like) surface magnetic fields and temperatures neither ions nor electrons would not be trapped, implying that the vacuum gaps envisions by RS75 would not form for normal pulsars.  If $\Omega \cdot B > 1$, electrons would be lifted from the polar caps but their flux is limited to the GJ flux, $\dot n_{\rm GJ} = \rho_{\rm GJ} A_{\rm PC} c$, so-called 
`space-charge limited flow', where $A_{\rm PC}$ is the polar cap area.  In this case, the steady flow of electrons supplies a fraction of $\rho_{\rm GJ}$ that decreases with distance above the neutron star, causing an increasing $E_\parallel$ (\cite[Arons \& Scharlemann 1979)].  The accelerating electrons radiate CR $\gamma$-rays that pair produce and screen the $E_\parallel$ above a pair formation front (PFF) by accelerating a small fraction of positrons back to the star.  Near the last open field line at the edge of the polar cap, where $E_\parallel \rightarrow 0$, pairs aren't produced since the elections cannot accelerate rapidly enough, forming a `slot gap'.  The limitation of the voltage to $V_0 \sim 10^{13}$ eV by pairs turns out to be very insensitive to $P$ and $\dot P$, so that the expected $\gamma$-ray luminosity $L_{\gamma} = \dot n_{\rm GJ}\,V_0 \propto \dot E_{\rm rot}^{1/2}$ (Harding 1981), a trend that is now observed for young $\gamma$-ray pulsars (\cite[Abdo et al. 2013)].  

\cite{DH82} simulated the pairs cascade above the PFF as the primary electrons continue to emit CR photons that produce pairs that produce synchrotron radiation (SR) photons that produce more pairs, estimating that the total cascade multiplicity, the number of pairs per primary electron, could reach $10^3 - 10^4$.  They also computed the spectrum of $\gamma$-rays that would be produced by the pair cascades (\cite[Daugherty \& Harding 1982, 1996)].  Two major developments for polar cap models were the addition of general relativistic effects that increase the $E_\parallel$ through frame-dragging (\cite[Muslimov \& Tsygan 1992)], and the finding that pairs can also be produced by resonant inverse Compton scattering (ICS) of surface thermal X-rays (\cite[Zhang et al. 1996]{Zhang1996}). Making use of these ideas, \cite[Harding \& Muslimov (2001,2002)]{HM01} modeled CR and ICS PFFs, pair death lines and polar cap heating luminosities.  

There were also ideas for producing (particularly high-energy) radiation in the outer magnetosphere.  \cite{Morini83} discovered that relativistic aberration and time delays of emission beamed along the last open field lines of a magnetic dipole would cancel the phase differences caused by the field-line curvature, bunching radiation into a narrow phase range or `caustic'.  Such caustics could produce the observed high-energy pulses, an idea also proposed by \cite{Smith1986}.   These models however consider only the geometry of the emission.  \cite{CHR85} originated the outer gap model in which a vacuum gap could exist between the null charge surface and the LC if particles of one sign flow out along open field lines but there are no charges below to replace them.  This gap accelerates charges to produce CR, but near the LC pairs are created by the photon-photon process, using thermal X-rays from the neutron star surface, since the field is too low for magnetic pair production.  The 3D geometry of the outer gap was modeled by \cite{RY95}, using the \cite{Deutsch1955} retarded vacuum dipole magnetic field, producing $\gamma$-ray light curves with caustic peaks from a single magnetic pole.  They predicted a correlation between the separation of the light curve peaks and the phase lag of the first peak with respect to the radio peak, now seen in the Fermi pulsar light curves (\cite[Abdo et al. 2013]{Abdo2013}).  \cite{HS01} pointed out that the previous outer gap $E_\parallel$ estimates were inconsistent since the screening by pairs had not been considered.  They modeled self-consistent outer gaps, concluding that the gap cannot provide the observed $\gamma$-ray luminosity by itself but required an incoming current, extending the gap boundary toward the neutron star.  

Another idea for the geometry of light curves from the outer magnetosphere was the `two-pole caustic' (TPC) model (\cite[Dyks \& Rudak 2003]{DyksRudak2003}), that assumed radiation along the last open field lines from the neutron star surface to near the LC.  The resulting light curve  have emission peaks from opposite poles.  \cite{MH03}, 2004; \cite[Harding et al. 2008]{Harding2008}) revived the slot gap of \cite{AS79}, adding frame dragging, to model the high energy emission by the accelerated particles.  The geometry is the same as that of the TPC model and the light curves look similar.  The first model for emission from the current sheet outside the LC was proposed by \cite{Petri2005}, who modeled SR from particles in the striped wind.  \cite{Petri2012} showed that such emission could produce caustic peaks (see also \cite[Bai \& Spitkovsky 2010]{BS2010}), but the radio phase lag was larger than for TPC model and not in agreement with the Fermi data.  \cite{US2014} also proposed that the GeV gamma-ray emission was SR from the current sheet, where particles were energized by magnetic reconnection.
 
\section{Magnetosphere Models}

Although the concept of a force-free (FF) magnetosphere filled with plasma was envisioned by \cite{GoldreichJulian1969}, the equation describing the fields and currents of an aligned FF magnetosphere, the `Pulsar Equation',  was introduced by \cite{Michel1973}.  He was not able to derive a solution for a rotating magnetic dipole, he was able to derive a solution for the split monopole which approximately describes the fields at large distances from the neutron star.  He also pointed out that for a dipole field, a current sheet would form along the rotation equator as the field lines from opposite poles merge beyond the LC (\cite[Michel 1975]{Michel1975}).  \cite{Beskin1983} obtained approximate solutions to the Pulsar Equation for oblique rotators and further noted that there is a contribution from the electric current to the Poynting flux which makes up the entire Poynting flux for the aligned rotator.  Magnetic reconnection of field lines in the striped wind of the current sheet was proposed by \cite{Coroniti1990} and \cite{Bogovalov1999} provided an analytic solution for the current sheet of an oblique rotator.

How the FF magnetosphere fills with charge was an unanswered question.  A numerical N-body experiment allowing either sign of charge to be pulled from the neutron star and accelerated by the vacuum $E_\parallel$ for an aligned rotator showed that two separated regions of static and opposite charge form above the polar caps (dome) and along the equator (torus) (\cite[Krauss-Polsdorff \& Michel 1985]{KM85}).  This configuration is a dead pulsar with no currents or acceleration.  However, a number of later studies using particle-in-cell (PIC) simulations found that the torus develops a diocotron instability that may allow the magnetosphere to fill with charge (\cite[Petri et al. 2002]{Petri2002}).

The first numerical solution for the axisymmetric FF magnetosphere was obtained by \cite{CKF1999} by requiring that ${\bf E\cdot B} = 0$ everywhere and iterating the current distribution to find a static configuration, showing that a current sheet forms along the spin equator and the poloidal field lines beyond the LC straighten into a monopole.   The current density distribution across the polar cap in this solution shows that the current flows out along field lines near the magnetic poles and returns along the current sheet and separatrix (along the last open field line) to the star (see also \cite[Timokhin 2006]{Timokhin2006}).  Numerical solutions for the oblique FF magnetospheres were derived by solving the time-dependent Maxwell's Equations (\cite[Spitkovsky 2006]{Spit2006}, \cite[Kalapotharakos \& Contopoulos 2009]{KC2009}), showing that the return current region becomes more distributed and axisymmetric with increasing inclination angle.  \cite{Timokhin2010} and \cite{TA2013} found that polar cap cascades, previously modeled in the steady-state limit, cannot supply the FF currents unless the cascades were non-steady.  The non-steady cascades have higher maximum voltage than steady cascades and can therefore produce larger pair multiplicities above $10^5$ (\cite[Timokhin \& Harding 2015]{Timokhin2015}).  

\section{Toward Self-Consistency}

The FF models however do not describe real pulsars since there is no $E_\parallel$ acceleration and no radiation.  If the FF condition ${\bf E\cdot B} = 0$ is relaxed, dissipative magnetosphere solutions can be found for different values of a macroscopic conductivity $\sigma$ (\cite[Kalapotharakos et al. 2012]{Kala2012}, \cite[Li et al. 2012]{Li2012}).  These solutions span the range between vacuum and FF magnetospheres and showed self-consistent regions of $E_\parallel$.  Using models with infinite $\sigma$ (FF) inside the LC and finite $\sigma$ outside the LC, \cite{Kala2014} found that the particle acceleration and radiation patterns matched the characteristics of the Fermi light curves and phase-resolved spectra (\cite[Brambilla et al. 2015]{Brambilla2015}) and that $\sigma$ should increase with $\dot E$.  However, the dissipative models are not completely self-consistent since the microphysics that creates the $\sigma$ distribution is not specified.  

To fully compute the self-consistent feedback between particle motions and fields, PIC simulations are needed.  The first PIC simulation of a pulsar magnetosphere were performed by \cite{Phil2014} and \cite{Phil2015}, using a Cartesian grid and injecting pair plasma ($e^+-e^-$ pairs) throughout the computational domain.  
\cite{Cerutti2015} used a spherical 2D PIC code to show that injecting enough pair plasma only above the neutron star surface could produce a near-FF solution for an aligned rotator.  Simulations requiring arbitrary thresholds on particle energies for pair injection found that pair production must occur in the current sheet as well as at the polar caps to create a FF solution (\cite[Chen \& Belodorodov 2014]{CB2014}, \cite[Philippov et al. 2015]{Phil2015}).  Using a 3D Cartesian PIC code, Kalapotharakos et al. (2017) found that if larger pair multiplicity is injected everywhere or  from the neutron star surface (\cite[Brambilla et al. 2017]{Brambilla2017}), FF magnetospheres can be formed at all inclination angles without the need for pair production in the current sheet.  For increasing injection rates, the $E_\parallel$ is confined more to the current sheet, where the highest energy particles are found.

High-energy radiation and light curves using PIC solutions were modeled by \cite{Cerutti2016}, finding that accelerated positrons in the current sheet produce this radiation.  \cite{Kala2017} modeled the high-energy emission from a 3D PIC code scaling the PIC energies, that are limited to $\gamma \sim 10^3$, up to those in real pulsars ($\gamma \sim 10^7$) and compared with Fermi data.  They found that the Fermi pulsar spectral cutoff energies and luminosities require pair injection rates that increase with spin-down rate, providing a physical link between microscopic injection rate and macroscopic $\sigma$.  

Although pulsar emission physics has made amazing progress over the last 50 years, with the primary location of the high-energy emission migrating from polar caps to the current sheet outside the LC, even the latest PIC models are not yet fully self-consistent.  The present PIC codes cannot resolve the small scales necessary to simulate the microphysics of pair creation and further progress may require hybrid codes.  So there remain more challenges for the next 50 years.

\end{document}